\documentclass[letterpaper,12pt]{article}
\usepackage{graphicx}
\usepackage{color,amssymb,enumerate,float,amsmath}
\usepackage{tikz}
\usetikzlibrary{matrix}
\usepackage{caption}
\usepackage{subcaption}
\usepackage{ifpdf}
\ifpdf
	\usepackage[pdftex,unicode,implicit]{hyperref}

	\hypersetup{
  	pdftitle     = {}, 
  	pdfkeywords  = {},
  	pdfauthor    = {},
  	pdfcreator   = {pdf\LaTeXe\ with package \flqq hyperref\frqq},
  	pdfproducer  = {pdf\LaTeXe\ with package \flqq hyperref\frqq},
  	pdfpagemode  = UseNone,  
  	pdffitwindow = true,  
  	unicode      = true,
  	plainpages   = true,
  	colorlinks   = true,  
  	citecolor    = black,  
  	urlcolor     = blue, 
  	linkcolor    = black
	}

\else

  \usepackage[dvips]{graphicx}

\fi 

\makeatletter
\@addtoreset{equation}{section}
\makeatother

\begin{document}
	
\thispagestyle{empty}

\begin{center}
{\bf \LARGE Casimir effect in 2+1 Ho\v rava gravity}
\vspace*{15mm}

{\large Claudio Bórquez}$^{1,a}$
{\large and Byron Droguett}$^{2,b}$
\vspace{3ex}

$^1${\it Department of Physics, Universidad de Antofagasta, 1240000 Antofagasta, Chile.
}
$^2${\it Departamento de Ciencias B\'asicas, Facultad de Ciencias, Universidad Santo Tom\'as, Sede Arica, Chile.}

\vspace{3ex}

$^a${\tt cl.borquezg@gmail.com,}\hspace{.5em}
$^b${\tt byrondroguett@santotomas.cl}

\vspace*{15mm}
{\bf Abstract}
\begin{quotation}{\small\noindent

We study the Casimir effect of a membrane embedded in $2+1$ dimensions flat cone generated by a massive particle located at the origin of the coordinate system. The flat cone is an exact solution of the nonprojectable Ho\v rava theory, similar to general relativity. We consider a scalar field satisfying Dirichlet boundary conditions, and regularize the spectrum using the $\zeta$--function technique. In addition, we include the effects of temperature in our analysis. Our results show that the Casimir force depends on three factors: the anisotropic scaling $z$, the mass of the point particle, and the temperature.
}
\end{quotation}
\end{center}

\thispagestyle{empty}

\newpage

\section{Introduction}
The Casimir effect is a physical manifestation of the quantum fluctuations of empty space. Casimir predicted the force per unit area between two uncharged, parallel, and conducting plates is attractive in $3+1$ dimensions \cite{Casimir:1948dh}. Other configurations also have been studied, such as two spheres, which exhibit the opposite sign for the Casimir force \cite{Boyer:1968uf}. Theoretical results for the Casimir effect generally agree with laboratory measurements \cite{Lamoreaux:1996wh, Bressi:2002fr}. The effect depends on the geometry  of the boundaries and the structure of the manifold considered \cite{Bordag:2001qi}. The temperature also  plays a crucial role in the results. It is often helpful to approach the problem within the framework of quantum field theory, which it allows us to associate operators that satisfy boundary conditions and incorporate temperature through an effective action \cite{Kirsten:2010zp}. Then, the effective path integral can be expressed in terms of a $\zeta$--function, which allows us to calculate the spectral sum over all eigenvalues associated to some operator. In this research, we focus on Dirichlet boundary conditions and finite temperature in a $2+1$ dimensional manifold.

Our aim is to study the Casimir effect in theories where the Lorentz symmetry is broken, such as Ho\v rava-Lifshitz-like theories. There have been several studies on the violation of Lorentz symmetry in this context \cite{Anselmi:2008bq,Anselmi:2008bs}. Several cases with anisotropic behavior have been considered in the literature, such as the extensions of Klein-Gordon and fermionic fields \cite{Ferrari:2010dj,MoralesUlion:2015tve,Muniz:2014dga,daSilva:2019iwn}. Other studies on Lorentz violation have included terms in the Lagrangian with a preferred direction \cite{Cruz:2017kfo,Erdas:2020ilo}. Additionally, finite temperature problems in quantum field theory have been analyzed \cite{Cruz:2018bqt,Erdas:2021xvv,Cheng:2022mwd}.

The framework in which we calculate the Casimir effect is in the context of the Ho\v rava gravity theory \cite{Horava:2009uw, Horava:2008ih}. It is a proposal for quantizing the ultraviolet sector of general relativity that involves Lorentz symmetry breaking through  an anisotropic scaling between space and time. The theory has a preferred foliation with absolute physical meaning, and the diffeomorphisms that preserve this foliation (FDiff) ensure the existence of a vector invariant that depends on the lapse function (in the nonprojectable case) \cite{Blas:2009qj}. In a previous study in $2+1$ dimensions, an exact solution was found, which geometrically represents a cone with a deficit or excess angle, with a massive point particle localized at the origin \cite{Bellorin:2019zho}. This solution is similar to one found in general relativity \cite{Deser:1983tn}, except for a dimensionless coupling constant. It is a solution that helped to define the condition of asymptotic flatness exactly in the same way as $2+1$ general relativity  \cite{Ashtekar:1993ds}. Motivated by this solution, we place a  finite membrane on the cone which satisfies  Dirichlet boundary conditions. In this investigation, we consider the case of a scalar field, which generates the quantum fluctuations of the vacuum. We examine how the mass of the particle located at the origin affects the Casimir energy and force. Finally, we make an analytic extension to explore the effects of the finite temperature on Casimir force.

This paper is organized as follows. In section 2, we introduce  Ho\v rava gravity and explain as was calculated the exact  solution in $2+1$ dimensions. In section 3, we calculate the Casimir force of a finite membrane at zero temperature. In  section 4, we include finite temperature in the calculation of the Casimir force. Finally, in section 5, we present our conclusions.


\section{Ho\v rava gravity}

The Ho\v rava theory \cite{Horava:2009uw} is a proposal to complete the ultraviolet regime of general relativity using quantum field theories techniques, making it unitary and power-counting renormalizable. This theory has a symmetry under anisotropic scaling of the coordinates, which is given by
\begin{equation}
	[t]=-z\,,
 \qquad
 [x^i]=-1
 \,.
\end{equation}
The price to pay for this theory is the Lorentz symmetry breaking in the ultraviolet, which arises as an accidental symmetry at large scales.
The foliation of spacetime has absolute physical meaning. The Arnowitt-Deser-Misner variables $N$, $N_i$, and $g_{ij}$ are used to describe the gravitational dynamics on the foliation. As a result, the general diffeomorphisms of general relativity are broken, and it is possible to introduce higher spatial derivatives into the Lagrangian while it keeps the second order time derivative under control. The symmetry group characteristic of the theory is given by the foliation-preserving diffeomorphisms (FDiff). The infinitesimal transformations are
\begin{equation}
	\delta t=f(t)
 \,,
 \qquad 
 \delta x^{i}
 =
 \zeta^{i}(t,\vec{x}) 
 \,.
\label{fdiffcoord}
\end{equation}
These induce the transformations on the fields
\begin{eqnarray}
	\delta N 
 &=& 
	\zeta^{k} \partial_{k} N 
 + f \dot{N} 
 + \dot{f}N
 \,, 
	\label{deltaN}
	\\ 
	\delta N_{i}
 &=& 
	\zeta^{k} \partial_{k} N_{i} 
 + N_{k} \partial_{i} \zeta^{k} 
 + \dot{\zeta}^{j} g_{ij} 
 + f \dot{N}_{i} 
 + \dot{f}N_{i} \,,
	\\
	\delta g_{ij} 
 &=&
 \zeta^{k} \partial_{k} g_{ij} 
 + 2g_{k(i}\partial_{j)} \zeta^{k} 
 + f \dot{g}_{ij} 
 \,.
\end{eqnarray}
The transformation of the lapse function (\ref{deltaN}) guarantees the existence of two versions: a projectable version (the lapse function depends only on time) and a nonprojectable version (the lapse function depends on time and space). In this paper, we work with the nonprojectable version since it is closer to general relativity.

A consequence is the theory propagates an instantaneous scalar mode in all dimensions, yielding a nontopological theory in $2+1$ dimensions. This scalar mode is responsible for the gravitational interaction in this dimension, making it an excellent laboratory for studying the perturbative quantization and black hole solutions in the UV complete theory.

In a previous work \cite{Bellorin:2019zho}  we coupled a relativistic particle to the Ho\v rava action in $2+1$ dimension considering the infrared terms in the potential 
\begin{eqnarray}
    \mathcal{V}
    &=& 
    -\beta R 
    -\alpha a_ka^k
    \,,
\end{eqnarray}
where the vector $a_k=\partial_k\ln(N)$ is invariant under FDiff \cite{Blas:2009qj}.

The combined system Ho\v{r}ava gravity-point particle in $2+1$  dimensions is given by the action 
\begin{equation}
	S = 
	\frac{1}{2\kappa} \int \,dt \,d^2x 
	\sqrt{g} N \left( K_{ij}K^{ij} - \lambda K^2 
	+ \beta R + \alpha a_k a^k \right) 
	- M \int dt \sqrt{ L } \,,
	\label{action}
\end{equation}
where 
\begin{eqnarray}
 &&
 K_{ij} = 
 \frac{1}{2N} 
 \left(
 \dot{g}_{ij} - 2 \nabla_{(i} N_{j)}
 \right) 
 \,,
 \label{extrinsiccurvature}
 \\ && 
 L = 
 (N^{2} - N_{k}N^{k}) 
 \left(
 \dot{q}^0 
 \right)^2 
 - 2 N_k \dot{q}^0 \dot{q}^k - g_{kl} \dot{q}^k \dot{q}^l
 \,.
 \label{L}
\end{eqnarray} 
The coefficient $M$ is the mass of the particle, $\kappa$ and $\lambda$ are coupling constants. The tensor (\ref{extrinsiccurvature}) is the extrinsic curvature. $L$ is the squared line element of the particle evaluated on the background of the ADM variables, and these variables are evaluated at the position of the particle in $L$. The mechanic of the particle is characterized by the embedding fields $q^{0} = q^{0}(t)$ and $q^{i} = q^{i}(t)$, which define the position of the particle in the foliation.

If we consider the particle at rest in the origin coordinate system, and all the fields are considered static, then it is possible to obtain an exact solution similar to topological general relativity, except for a dimensionless constant $\beta$. This exact solution has the form
\begin{eqnarray}
	ds^{2} 
 &=& 
	r^{-\frac{ \kappa M}{\pi\beta}}
 (
 dr^2 
 + r^2 d\theta^2
 ) 
 \,.
	\label{solexacta}
\end{eqnarray}
This solution is not trivial due to the complexity of the equations of motion, and because the nonprojectable version has second-class constraints, unlike general relativity. The dimensionless constant $\beta$ is fixed to $1$ by general symmetry in general relativity. From a geometric point of view, it is convenient to make the following coordinate change
\begin{equation}
	\rho =
 \frac{1}{\gamma} r^{\gamma} 
 \,,
	\quad
	\theta\,' = 
 \gamma \theta 
 \,,
	\quad
	\gamma \equiv 1
 - \frac{ \kappa M}{2 \pi \beta}
 \,,
	\label{cambiodevariable}
\end{equation}
this leads to a flat cone solution with a deficit or excess angle
\begin{equation}
	ds^2 =
 d\rho^2 
 + \rho^2 d{\theta\,'}^2
 \,.
 \label{solplana}
\end{equation}
For the case where $\gamma > 0$, the solution represents a flat cone with a deficit or excess angle, with the particle localized at the origin. The domain of this solution is $\rho \in [0, \infty)$ and $\theta\,' \in [0, 2\pi\gamma]$. When $M = 0$, the geometry is globally plane. In the case where $\gamma < 0$, there is no physical solution because distances are not well defined. The physical solution $\gamma > 0$ motivated us to define the asymptotic flatness conditions of the Ho\v rava theory in $2+1$ dimensions, similar to general relativity \cite{Ashtekar:1993ds}. The solution is completely global  and the Newtonian force is identically zero; the cone is affected globally. In addition, the perturbative wave equation for scalar mode does not determine the sign of the coupling constant associated to the Ricci scalar \cite{Bellorin:2019zho}. Therefore, it is important to research other physical quantities in order to determine this value.
Motivated by this exact solution in $2+1$ dimensions, we calculate the Casimir energy and force of a membrane localized on the flat cone which satisfies Dirichlet boundary conditions.


\section{Casimir effect in Ho\v rava-Lifshitz theory}

The modified Klein-Gordon Lagrangian for a scalar field invariant under the anisotropic scaling is given by 
\begin{eqnarray}
    S_{\phi}
    &=&
    \frac{1}{2}\int\,dt\,d^dx\sqrt{g}
    \left(
    \partial_t\phi\partial_t\phi
    -l^{2(z-1)}\partial_{i1}\partial_{i2}\cdots\partial_{iz}\phi\partial^{i1}\partial^{i2}\cdots\partial^{iz}\phi
    \right)\,,
\end{eqnarray}
and the equation of motion for the scalar field is
\begin{equation}
(\partial_t^2+(-1)^zl^{2(z-1)}\Delta^z)\phi=0\,,
\end{equation}
where $\Delta=g^{ij}\nabla_i\nabla_j$ and the parameter $l$ has dimension of the inverse of mass. 

The problem we aim to solve is the embedding of a membrane on a flat cone with a deficit or excess angle in $2+1$ dimensions. We solve the eigenvalues problem for a scalar field with arbitrary $z$ value satisfying Dirichlet boundary conditions
\begin{equation}
    \mathcal{P}\phi=(-1)^zl^{2(z-1)}\Delta^{z}\phi=
    \omega_z\phi\,, 
    \qquad
    \phi(\theta_1)=
    \phi(\theta_2)=0
    \,,
    \qquad 
    \phi(R_1)=
    \phi(R_2)=0\,.
    \label{boundary}
\end{equation}
First, we consider the case $z=1$ in order to obtain its eigenvalues, which will help us  to find the eigenvalues form for an arbitrary $z$ value. Using the exact solution (\ref{solexacta}), the partial differential  equation is given by
\begin{equation}
-\Delta\phi=
- r^{\mu}\left(
\dfrac{\partial^{2}\phi}{\partial r^{2}}
+\frac{1}{r}\dfrac{\partial\phi}{\partial r}
+\frac{1}{r^{2}}\dfrac{\partial^{2}\phi}{\partial \theta^{2}}\right)
=
\omega_1\phi
\,,
\end{equation}
where $\mu=-\frac{ \kappa M}{\pi\beta}$. By using  separable variables $\phi=R(r)\Theta(\theta)$, we obtain two equations given by
\begin{eqnarray}
\Theta''(\theta)
+k^{2}\Theta(\theta)
&=&0\,,
\label{G}
\\
r^{2}R''(r)
+rR'(r)
+\left(
\omega_1 r^{2-\mu}
-k^{2}
\right)R
&=&0\,,
\label{RR}
\end{eqnarray}
To solve Eq. (\ref{RR}), we make the following change of variables:
\begin{equation}
x=
\sigma r^{\gamma}
\,,
\qquad
\sigma=
\frac{\sqrt{\omega_1}}{\gamma}
\,,
\end{equation}
where $\gamma=1-\mu/2$. This results in the standard Bessel differential equation, whose solution can be expressed in terms of Bessel functions of the first and second kind
\begin{eqnarray}
    R(r)
    =
C_1J_p\left(
\frac{\sqrt{\omega_1}}{\gamma} r^\gamma
\right)
    +C_2Y_p\left(
    \frac{\sqrt{\omega_1}}{\gamma} r^\gamma
    \right)
    \,,
\end{eqnarray}
with $p=k/\gamma$. For simplicity, we consider asymptotic behavior of the Bessel functions
\begin{eqnarray}
J_p\left(
\frac{\sqrt{\omega_1}}{\gamma} r^\gamma
\right)
&\sim&
\sqrt{\frac{2\gamma}{\pi \sqrt{\omega_1}r^{\gamma}}}\cos 
\left(
\frac{\sqrt{\omega_1}r^{\gamma}}{\gamma}
-p\pi/2
-\pi/4
\right)
\,,
\label{asymmptotic}
\\
Y_p
\left(
\frac{\sqrt{\omega_1}}{\gamma} r^\gamma
\right)
&\sim& 
\sqrt{\frac{2\gamma}{\pi \sqrt{\omega_1}r^{\gamma}}}\sin 
\left(
\frac{\sqrt{\omega_1}r^{\gamma}}{\gamma}
-p\pi/2
-\pi/4
\right)
\,.
\label{asymptoticc}
\end{eqnarray}
Therefore, the Dirichlet boundary conditions determine the form of the eigenvalues 
\begin{equation}
    \omega_1=
    \left(
    \frac{n\pi\gamma}{R_{2}^{\gamma}
    -R_{1}^{\gamma}}
    \right)^{2}
    \,,
    \qquad 
    n\in\mathbb{N}
    \,.
\end{equation}
With this result, we can generalize the eigenvalues solution of the operator $\mathcal{P}$ in Eq. (\ref{boundary}) for an arbitrary $z$ value. Thus, the eigenvalues are given by
\begin{equation}\label{omegaz}
    \omega_z=
    l^{2(z-1)}
    \left(
    \frac{n\pi\gamma}{R_2^\gamma-R_1^\gamma}
    \right)^{2z}
    \,.
\end{equation}
The vacuum expectation value of the energy must be regularized using the $\zeta$--function associated to the spatial operator $\mathcal{P}$
\begin{equation}
\zeta_{\mathcal{P}}(s)
=
l^{-2(z-1)s}
\sum_{n\in\mathbb{N}}
\left(
\frac{n\pi\gamma}{R_{2}^{\gamma}
-R_{1}^{\gamma}}
\right)^{-2sz}
\label{zetasinTz}
\,.
\end{equation}
Therefore the Casimir energy is given by
\begin{equation}
    E_{C}=\frac{1}{2}\zeta_{\mathcal{P}}\left(-1/2\right)=
  \frac{1}{2} l^{z-1} 
  \left(
  \frac{\pi\gamma}{R_{2}^{\gamma}
-R_{1}^{\gamma}}
\right)^{z}\zeta_R(-z)
\,,
\end{equation}
where $\zeta_{R}$ is the Riemann $\zeta$--function. If we derive with respect to the separation of the membrane $d_{\gamma}=\frac{1}{\gamma}(R_{2}^{\gamma}
-R_{1}^{\gamma})$, we get the Casimir force
\begin{eqnarray}\label{FCz1}
    F_{C} &=&
    \frac{1}{2}z\pi^{z}l^{z-1}\left(\frac{\gamma}{R_{2}^{\gamma}-R_{1}^{\gamma}}\right)^{z+1}\zeta_{R}(-z)\,.
\end{eqnarray}
The Riemann $\zeta$--function is a global factor of the energy and force. Here, two cases for integer values of $z$ can occur: first, if $z$ is an even number, the energy and force are equal to zero. Second, if $z$ is an odd number, the Casimir force can be either attractive or repulsive. When the separation $d_\gamma$ tends to infinity the Casimir force tends to zero for all $\gamma$.

\begin{figure}[H]
    \begin{minipage}[b]{0.5\linewidth}
    \centering
    \includegraphics[width=.8\linewidth]{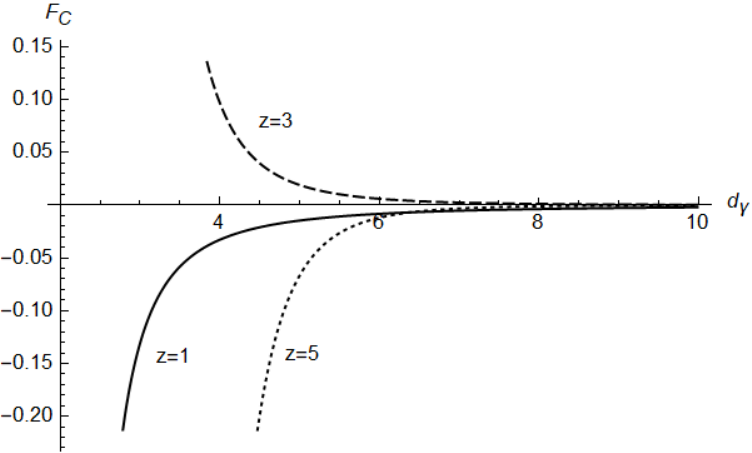} 
    \caption*{(a)}
    \vspace{4ex}
  \end{minipage}
    \begin{minipage}[b]{0.5\linewidth}
    \centering
    \includegraphics[width=.8\linewidth]{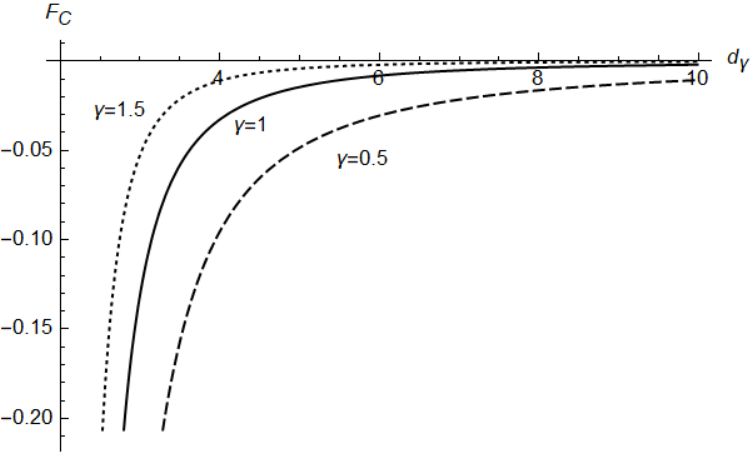}
    \caption*{(b)}
    \vspace{4ex}
  \end{minipage}
  \caption{(a): Casimir force versus separation $d_{\gamma}$, considering $\gamma=1$. (b): Casimir force versus separation $d_{\gamma}$, considering $z=1$.}
  \label{fig1}
  \end{figure}

In the Fig.\ref{fig1} (a), we consider the globally flat case $\gamma=1$. The force as a function of distance decays as expected. The unusual aspect is the change in the orientation of the force each time we fixed different $z$ values, such as it is presented in \cite{MoralesUlion:2015tve}. This is due directly to the Riemann $\zeta$--function. For the case where $z$ takes on even values, the force is zero. In the Fig.\ref{fig1} (b), the force tends to zero for different values of the parameter $\gamma$. Here, we note that the magnitude of the force is modified, that is, the force will be  stronger (for a cone with deficit angle) or weaker (for a cone with excess angle) depending on the $\gamma$ factor.

\section{Casimir force at finite temperature}

In order to study the effect of temperature, it is convenient to use the path integral approach considering imaginary time, which is associated to finite temperature in the system. The path integral for the scalar field is given by
\begin{equation}
    Z=
    \int\mathcal{D}\phi\exp{(S(\phi))}\,.
\end{equation}
The effective action associated to operator $\mathcal{O}$ is 
\begin{equation}
    \Gamma=
    -\ln(Z)=
    \frac{1}{2}\ln\det[
    (
    -\partial^{2}_{\tau}
    +\mathcal{P}
    )/\epsilon
    ]=
    \frac{1}{2}\ln\det
    [
    \mathcal{O}/\epsilon
    ]\,,
\end{equation}
where $\epsilon$ is an arbitrary parameter with  mass dimension, introduced to render
the $\zeta$--function dimensionless. Eventually, the $\zeta$--function will be independent of this parameter $\epsilon$ hence we say $\epsilon=1$ for simplicity. Therefore, the Casimir energy is given by
\begin{eqnarray}
   E_{C}
   &=&
   \frac{\partial}{\partial\xi}\Gamma
   =
   \left.
   -\frac{1}{2}\frac{\partial}{\partial\xi}\left(\frac{d}{ds}\zeta_{\mathcal{O}}(s)
   \right)
   \right|_{s=0}
   \,,
\end{eqnarray}
where $\xi=1/T$ is the inverse of the  temperature.

The eigenvalues problem of the operator $\mathcal{O}$ is expressed by
\begin{eqnarray}
   (-
   \partial^{2}_{\tau}
   +(-1)^{z}l^{2(z-1)}\Delta^{z}
   )\phi
   =
   \omega\phi\,,
\end{eqnarray}
where $\tau\in\mathbb{C}$ and it has periodic border. We propose the  following solution to the scalar field
\begin{eqnarray}
   \phi_{m,n}(\tau,x^i)=
   \frac{1}{\xi}e^{\frac{2\pi im}{\xi}\tau}\varphi_{n}(x^i)
   \,,
\end{eqnarray}
where the eigenvalues associated to the time derivative are $\omega_{m}=\frac{2\pi m}{\xi}$, and the spatial eigenvalues come from the asymptotic behavior of the functions (\ref{asymmptotic}) and (\ref{asymptoticc}).
Then, the $\zeta$--function associated with the operator $\mathcal{O}$ is give by
\begin{eqnarray}
   \zeta_{\mathcal{O}}(s)
   &=&
   \sum_{m=-\infty}^{\infty}
   \sum_{n=1}^{\infty}
   \left[
   \left(
   \frac{2\pi m}{\xi}\right)^{2}
   +l^{2(z-1)}\left(\frac{n\pi\gamma}{R_{2}^{\gamma}
   -R_{1}^{\gamma}}
   \right)^{2z}
   \right]^{-s}\,.
\end{eqnarray}
We use the integral representation $\zeta$--function to rewrite the spectral function as
\begin{eqnarray}
   \zeta_{\mathcal{O}}(s)&=&
   \frac{1}{\Gamma
   \left(
   s
   \right)}
   \int_{0}^{\infty}dt\,
   t^{s-1}
   \sum_{m=-\infty}^{\infty}
   \sum_{n=1}^{\infty}
   \exp\left\lbrace
   -t\left[
   \left(
   \frac{2\pi m}{\xi}
   \right)^{2}
   +l^{2(z-1)}\left(
   \frac{n\pi\gamma}{R_{2}^{\gamma}
   -R_{1}^{\gamma}}
   \right)^{2z}
   \right]
   \right\rbrace\,.
   \nonumber\\
\end{eqnarray}
A suitable representation is obtained by using the Poisson resummation \cite{Kirsten:2010zp}
\begin{eqnarray}
 \zeta_{\mathcal{O}}(s)
 &=&
 \frac{\xi}{\sqrt{4\pi}}
 \frac{\Gamma\left(
 s-1/2
 \right)}
 {\Gamma\left(s\right)}
 \left[l^{2(z-1)}\left(\frac{\pi\gamma}{R_2^\gamma-R_1^\gamma}\right)^{2z}\right]^{1/2-s}
 \zeta_{R}
 \left(
 z(2s-1)
 \right)
 \nonumber\\
   &&
   +\frac{\xi}{\sqrt{\pi}\Gamma{(s)}}
   \sum_{n,m=1}^\infty
   \int_0^\infty\,dt
   \,t^{s-3/2}
   \exp
   \left(
   -\frac{\xi^2 m^2}{4t}
   -tl^{2(z-1)}
   \left(
   \frac{n\pi\gamma}{R_{2}^{\gamma}
   -R_{1}^{\gamma}}
   \right)^{2z}
   \right)\,.
   \nonumber\\
\end{eqnarray}
We can reduce the $\zeta$--function by introducing the modified Bessel function through the following change of variables
\begin{equation}
    y=
    l^{2(z-1)}\left(
    \frac{n\pi\gamma}{R_{2}^{\gamma}
    -R_{1}^{\gamma}}
    \right)^{2z}t
    \,,
    \qquad
    z=
    m\xi l^{z-1}
    \left(
    \frac{n\pi\gamma}{R_{2}^{\gamma}
    -R_{1}^{\gamma}}
    \right)^z
    \,,
\end{equation}
then, after performing the change, we have
\begin{eqnarray}
\zeta_{\mathcal{O}}(s)
&=&
 \frac{\xi}{\sqrt{4\pi}}
 \frac{\Gamma\left(
 s-1/2
 \right)}
 {\Gamma\left(s\right)}
 \left[l^{2(z-1)}\left(\frac{\pi\gamma}{R_2^\gamma-R_1^\gamma}\right)^{2z}\right]^{\frac{1}{2}-s}
 \zeta_{R}
 \left(
 z(2s-1)
 \right)
 \nonumber\\
   &&
   +\frac{2\xi}{\sqrt{\pi}\Gamma{(s)}}
   \sum_{n,m=1}^\infty \left\lbrace\left[\frac{2l^{z-1}}{m\xi}\left(\frac{n\pi\gamma}{R_{2}^{\gamma}-R_{1}^{\gamma}}\right)^{z}\right]^{\frac{1}{2}-s}K_{\frac{1}{2}-s}\left(\xi m l^{z-1}\left(\frac{n\pi\gamma}{R_{2}^{\gamma}-R_{1}^{\gamma}}\right)^{z}\right)\right\rbrace\,.
   \nonumber\\
\end{eqnarray}
Now, by expanding the derivative of the spectral function around $s=0$, we have
\begin{eqnarray}
    \left.\frac{d}{ds}\zeta_{\mathcal{O}}(s)\right|_{s=0}
    =
    -\xi l^{z-1} \left(
  \frac{\pi\gamma}{R_{2}^{\gamma}
-R_{1}^{\gamma}}
\right)^{z}\zeta_R(-z)
+ 2\sum_{m,n=1}^{\infty}\frac{1}{m}\exp\left(-\xi ml^{z-1}\left(
  \frac{n\pi\gamma}{R_{2}^{\gamma}
-R_{1}^{\gamma}}
\right)^{z}\right)\,.
\nonumber\\
\end{eqnarray}
The sum over $m$ can be explicitly performed using a geometric serie, then the Casimir energy is
\begin{eqnarray}\label{CasEner}
    E_{C}
    =
    l^{z-1} 
  \left(
  \frac{\pi\gamma}{R_{2}^{\gamma}
-R_{1}^{\gamma}}
\right)^{z}
\left[
\frac{1}{2}\zeta_R(-z)
+ \sum_{n=1}^{\infty}\frac{n^z}{\exp\left(\xi l^{z-1} 
  \left(
  \frac{n\pi\gamma}{R_{2}^{\gamma}
-R_{1}^{\gamma}}
\right)^{z}\right)-1}
\right]
\,.
\end{eqnarray}
In the case of $\gamma=1$, we recover the result for globally flat space. By taking the derivative with respect to the separation $d_{\gamma}$, we obtain the Casimir force
\begin{eqnarray}\label{FCz}
    F_{C} &=&
    z\pi^{z}l^{z-1}\left(\frac{\gamma}{R_{2}^{\gamma}
-R_{1}^{\gamma}}\right)^{z+1}
    \left\lbrace \frac{1}{2}\zeta_{R}(-z) + \sum_{n=1}^{\infty}\frac{n^z}{\exp\left[l^{z-1}\xi\left(\frac{n\pi\gamma}{R_{2}^{\gamma}
-R_{1}^{\gamma}}\right)^{z}\right]-1}\right.
    \nonumber
    \\
    &&-
    \left.\xi l^{z-1}\pi^{z}\left(\frac{\gamma}{R_{2}^{\gamma}
-R_{1}^{\gamma}}\right)^{z}\sum_{n=1}^{\infty}\frac{n^{2z}\exp\left[\xi l^{z-1}\left(\frac{n\pi\gamma}{R_{2}^{\gamma}
-R_{1}^{\gamma}}\right)^{z}\right]}{\left[\exp\left(\xi l^{z-1}\left(\frac{n\pi\gamma}{R_{2}^{\gamma}
-R_{1}^{\gamma}}\right)^{z}\right)-1\right]^{2}}\right\rbrace\,.
\end{eqnarray}

\begin{figure}[H]
    \begin{minipage}[b]{0.5\linewidth}
    \centering
    \includegraphics[width=.8\linewidth]{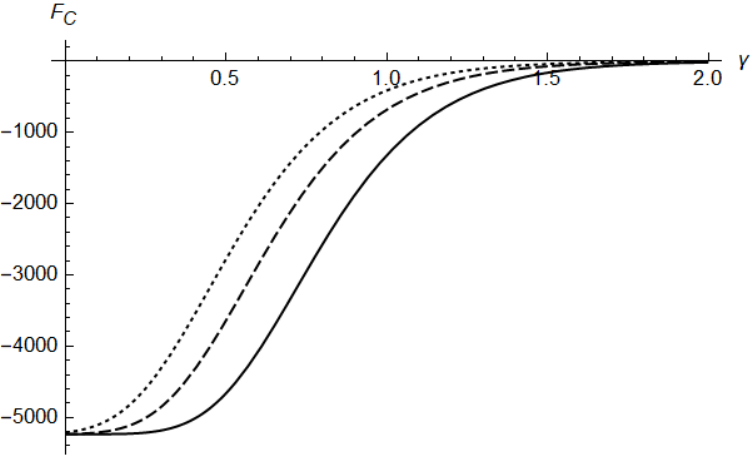} 
    \caption*{(a)}
    \vspace{4ex}
  \end{minipage}
    \begin{minipage}[b]{0.5\linewidth}
    \centering
    \includegraphics[width=.8\linewidth]{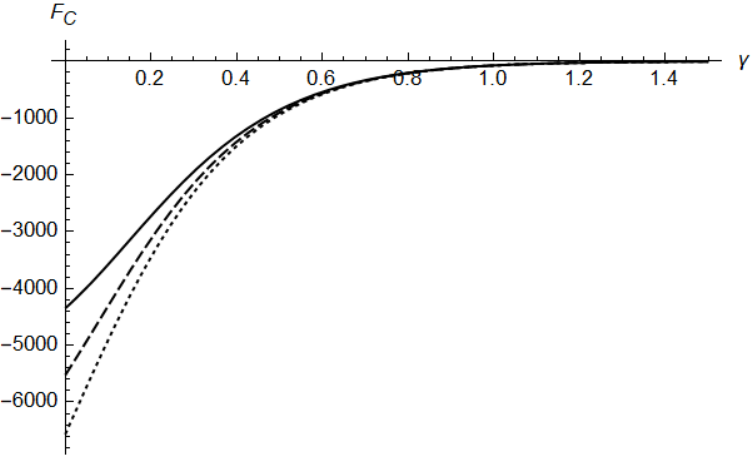}
    \caption*{(b)}
    \vspace{4ex}
  \end{minipage}
  \caption{(a): Casimir force versus  $\gamma$, $T=100$ and different separation $d_\gamma$. (b): Casimir force versus  $\gamma$, for $z=1$. The solid, dashed and dotted curves are $T=100 ,120, 140$, respectively.}
  \label{fig2}
  \end{figure}

\begin{figure}[h]
    \begin{minipage}[b]{0.5\linewidth}
    \centering
    \includegraphics[width=.8\linewidth]{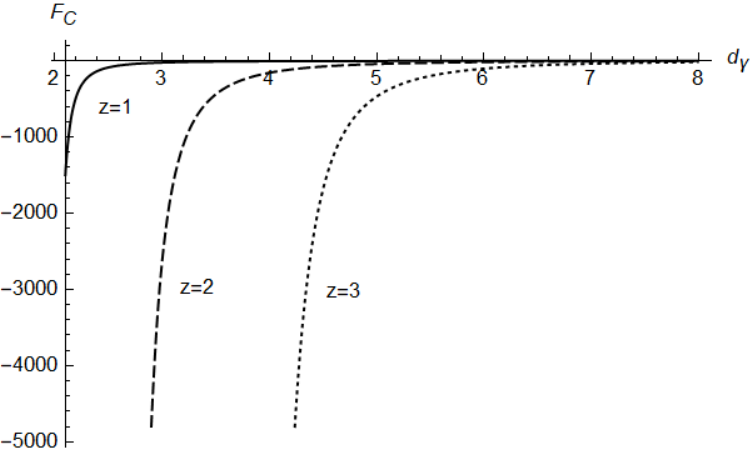} 
    \caption*{(a)}
    \vspace{4ex}
  \end{minipage}
    \begin{minipage}[b]{0.5\linewidth}
    \centering
    \includegraphics[width=.8\linewidth]{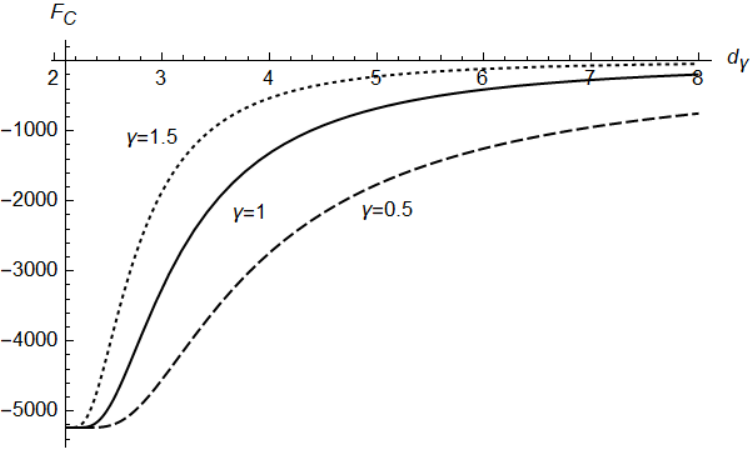}
    \caption*{(b)}
    \vspace{4ex}
  \end{minipage}
  \caption{(a): Casimir force versus  separation $d_\gamma$, $T=100$ and $\gamma=1$. (b): Casimir force versus separation $d_\gamma$,  $z=1$, and $T=100$ temperature.}
  \label{fig3}
  \end{figure}

 In Fig.\ref{fig2} (a) the behavior of the force as a function of $\gamma$ changes in magnitude for different separation distances by considering a fixed temperature. When the separation distance increases, the force decreases. The exponential factor $\gamma$ determines how the force decays. In Fig.\ref{fig2} (b) we show how the force varies with respect to $\gamma$, for different temperature values. At high temperatures, the magnitude of the force is greater, and the decay is faster than in the case of nontemperature. It is important to note that the temperature is independent of the factor $\gamma$.
In the Fig.\ref{fig3} (a), unlike what occurs in Fig. \ref{fig1} (a), we can see that for a certain finite temperature, and $\gamma=1$, the contribution of the sums in $n$ in Eq. (\ref{FCz}) are completely relevant. The orientation of the force for $z=1,2,3$ is the same, that is, there is no change in the sign of the force. In addition, for even values of $z$, there is a contribution to the Casimir force. In Fig. \ref{fig3} (b), we show how the force varies as a function of distance for three different values of $\gamma$ by considering a constant temperature and anisotropic factor $z=1$, such as in Fig. \ref{fig1} (b). The force decays to a finite value for different $\gamma$. This behavior was expected for temperatures other than zero. The magnitude of the force is stronger with a deficit angle.


\section{Conclusions}

We study the case of a membrane embedded in a flat cone with a deficit or excess angle generated by a massive particle located at the origin of the coordinate system. This cone is an exact solution of the Ho\v rava theory in $2+1$ dimensions. The vacuum fluctuations are described by a scalar field, which is expressed using the Ho\v rava-Lifshitz theory, and satisfies Dirichlet boundary conditions. We have derived the Casimir energy and force of this particular manifold at both zero and finite temperature.

At zero temperature, the force is multiplied by a global Riemann $\zeta$--function that depends on the anisotropic scaling factor $z$. If $z$ is even, the energy and force are zero. For the special case of $z=3$, the energy is positive, leading to a repulsive force. When temperature is taken into account, additional terms appear in the Casimir effect. If $z$ is even, the Casimir energy and force are nonzero, and specifically, for $z=1,2,3$, the force is attractive due to thermal influences. The factor $\gamma$ plays a significant role in the Casimir effect. When the cone has a deficit angle ($\gamma<1$), the magnitude of the force is greater than in the case of a flat space or excess angle ($\gamma\geq 1$).

The results are consistent with those found in the literature about the Casimir effect in Lorentz-violating theories: the Casimir effect depends on the anisotropic scaling $z$. Furthermore, the topology has a strong influence on the decay of the energy and force. Therefore, the Casimir effect in Ho\v rava-Lifshitz gravity theories will not only depend on boundary conditions, but also on the spatial configuration determined by the $\gamma$ parameter, that is, by the presence of a massive point particle.


\section*{Acknowledgements}
C.B.~is partially supported by Grant No. CONICYT PFCHA/DOCTORADO BECAS CHILE /2019 -- 21190960. C.B.~is a graduate student in the ``Doctorado en F\'isica Menci\'on F\'isica-Matem\'atica" Ph.D. program at the Universidad de Antofagasta.  


\end{document}